\documentclass[a4paper,fleqn,usenatbib]{mnras}

\usepackage{newtxtext,newtxmath}
\usepackage[T1]{fontenc}
\usepackage{ae,aecompl}

\usepackage{graphicx}	% Including figure files
\usepackage{amsmath}	% Advanced maths commands
\usepackage{epstopdf}
\graphicspath{{./Figures/}} % Specifies the directory where pictures are stored
\title[Follow-up of Gaia transiting planets]{On the follow-up efforts of long-period transiting planet candidates detected with Gaia astrometry}

\author[A. Sozzetti et al.]{
A. Sozzetti,$^{1}$\thanks{E-mail: alessandro.sozzetti@inaf.it}
P. Giacobbe,$^{1}$
M. G. Lattanzi,$^{1}$
and M. Pinamonti$^{1}$
\\
$^{1}$INAF - Osservatorio Astrofisico di Torino, Via Osservatorio 20, I-10025 Pino Torinese, Italy\\
}

\date{Accepted XXX. Received YYY; in original form ZZZ}

\pubyear{2016}

\begin{document}
\label{firstpage}
\pagerange{\pageref{firstpage}--\pageref{lastpage}}
\maketitle

\begin{abstract}

The class of transiting cold Jupiters, orbiting at $\gtrsim0.5-1.0$ au, is to-date underpopulated. Probing their atmospheric composition and physical characteristics is particularly valuable, as it allows for direct comparisons with the Solar System giant planets. We investigate some aspects of the synergy between Gaia astrometry and other ground-based and space-borne programs for detection and characterization of such companions. We carry out numerical simulations of Gaia observations of systems with one cold transiting gas giant, using Jovian planets around a sample of nearby low-mass stars as proxies. Using state-of-the-art orbit fitting tools, we gauge the potential of Gaia astrometry to predict the time of transit centre $T_c$ for the purpose of follow-up observations to verify that the companions are indeed transiting. Typical uncertainties on $T_c$ will be on the order of a few months, reduced to several weeks for high astrometric signal-to-noise ratios and periods shorter than $\sim3$ yr. We develop a framework for the combined analysis of Gaia astrometry and radial-velocity data from representative ground-based campaigns and show that combined orbital fits would allow to significantly reduce the transit windows to be searched for, down to about $\pm2$ weeks ($2-\sigma$ level) in the most favourable cases. These results are achievable with a moderate investment of observing time ($\sim0.5$ nights per candidate, $\sim50$ nights for the top 100 candidates), reinforcing the notion that Gaia astrometric detections of potentially transiting cold giant planets, starting with Data Release 4, will constitute a valuable sample worthy of synergistic follow-up efforts with a variety of techniques.

%This is a simple template for authors to write new MNRAS papers.
%The abstract should briefly describe the aims, methods, and main results of the paper.
%It should be a single paragraph not more than 250 words (200 words for Letters).
%No references should appear in the abstract.
\end{abstract}


%%%%%%%%%%%%%%%%%%%%%%%%%%%%%%%%%%%%%%%%%%%%%%%%%%

%%%%%%%%%%%%%%%%% BODY OF PAPER %%%%%%%%%%%%%%%%%%

\section{Introduction}

The class of long-period transiting giant planets, with orbital periods exceeding 1 yr, was first unveiled by the Kepler mission \cite{Borucki2010}. These objects are cold, with equilibrium temperatures $T_\mathrm{eq}\sim200$ K. They constitute a valuable sample for comparative studies of their atmospheric composition and physical properties with those of the outer planets of our Solar System. 
However, the combination of low geometric transit probability and low signal-to-noise ratio ($S/N$) of the events (due to their limited number) 
has thus far translated in a few tens of long-period systems identified in the full Kepler dataset based on two- or single-transit events (e.g., \citealt{Wang2015,Uehara2016,Foremanmackey2016,Beichman2018,Herman2019}). 
The comprehensive analysis of available data from the K2 mission has more recently allowed to uncover a large sample of mono-transit candidates (e.g., \citealt{Osborn2016,Lacourse2018}), some of which, with estimated durations of tens of hours, might indeed correspond to new 
detections of long-period gas giants beyond the snow line (one such instance being the case of EPIC248847494b reported by \citet{Giles2018}. More 
will come, with over 1000 single-transit candidates expected to be found by the TESS mission \citep{Villanueva2019,Kunimoto2022} and other significant numbers 
likely to be provided by the PLATO mission. 

Long-period gas giants producing single-transit events warrant follow-up efforts for accurate period and mass determination, in order to down-select 
the optimal target sample for atmospheric characterization with e.g. JWST \citep{Beichman2014}. Archival searches and carefully-planned photometric monitoring programs from the ground \citep{Cooke2018,Kovacs2019,Dholakia2020,Yao2019,Yao2021} and in space \citep{Cooke2019,Cooke2020} and ground-based radial-velocity (RV) work (e.g., \citealt{Hebrard2019,Gill2020,Dalba2020,Dalba2021a,Dalba2021b,Dalba2022,Ulmermoll2022}) are typically the channels used for the purpose. However, space-based high-precision astrometry with the Gaia mission \citep{Prusti2016,Vallenari2022}  also bears the potential for important contributions to this task. On the one hand, depending on target magnitude and distance, Gaia will provide useful mass upper limits or actual astrometric detections of long-period transiting planets, particularly in the regime of orbital separations $1-4$ au, for which Gaia achieves maximum sensitivity (e.g., \citealt{Lattanzi2000,Casertano2008,Sozzetti2014,Perryman2014}). Indeed, \citet{Holl2022} report close to edge-on orbital solutions for a  small sample of transiting systems identified by the Kepler and TESS mission, likely corresponding to the correct identification of the transiters as low-mass stars based on Gaia DR3 astrometry. On the other hand, Gaia might in fact be seen itself as a target provider for photometric and spectroscopic follow-up observations. Recent studies indicate that Gaia has the potential to identify astrometrically hundreds of giant-planet systems with $P\ge1$ yr having orbital inclinations compatible with a transit configuration \citep{Sozzetti2014}, some of which might be actually transiting \citep{Perryman2014}. In the sample of $\sim1900$ candidate substellar companions presented in \citet{Arenou2022}, 49 have orbital solutions with an inclination angle in the range [89,91] deg, i.e. compatible with a perfectly edge-on orbit, and the median period of these solutions is $\sim1.3$ yr. The population of potentially edge-on systems is likely underestimated, given Gaia's reduced sensitivity to such configurations extensively discussed in \citet{Arenou2022}. 

In this paper we investigate some aspects of the potential synergy between Gaia astrometry, Doppler measurements, and space-borne photometric time-series for improved characterization of long-period transiting giant-planet systems. 
%We initially explore the range of constraints on companion mass that Gaia might place on a representative 
%sample of known long-period Jupiter-sized companions uncovered by Kepler and K2 (Kepler-167e, Kepler-1654b, and EPIC248847494b). 
We focus our attention on the problem of combining Gaia astrometry and follow-up radial-velocity time-series for improved forecast of the time of transit center, in order to identify the preferred regime of orbital separations that might effectively be probed by space-based photometric observations with CHEOPS, TESS, and PLATO. For this task we utilize as reference sample the \citet{Lepine2011} all-sky catalog of bright, nearby M dwarfs, that maximizes the likelihood of high-precision orbit and mass determination with Gaia. All the findings described in this work can readily be scaled to other stellar and companion masses, and ranges of distance from the Sun.
In Section 2 we describe the adopted setup of the Gaia simulations, while in Section 3 we present the details of our analysis tools. The main results are presented in Section 4, followed by a brief summary and discussion.

\section{Simulation Scheme}

\subsection{Gaia astrometry}

\begin{figure*}
\centering
%$\begin{array}{cc}
\includegraphics[width=0.85\textwidth]{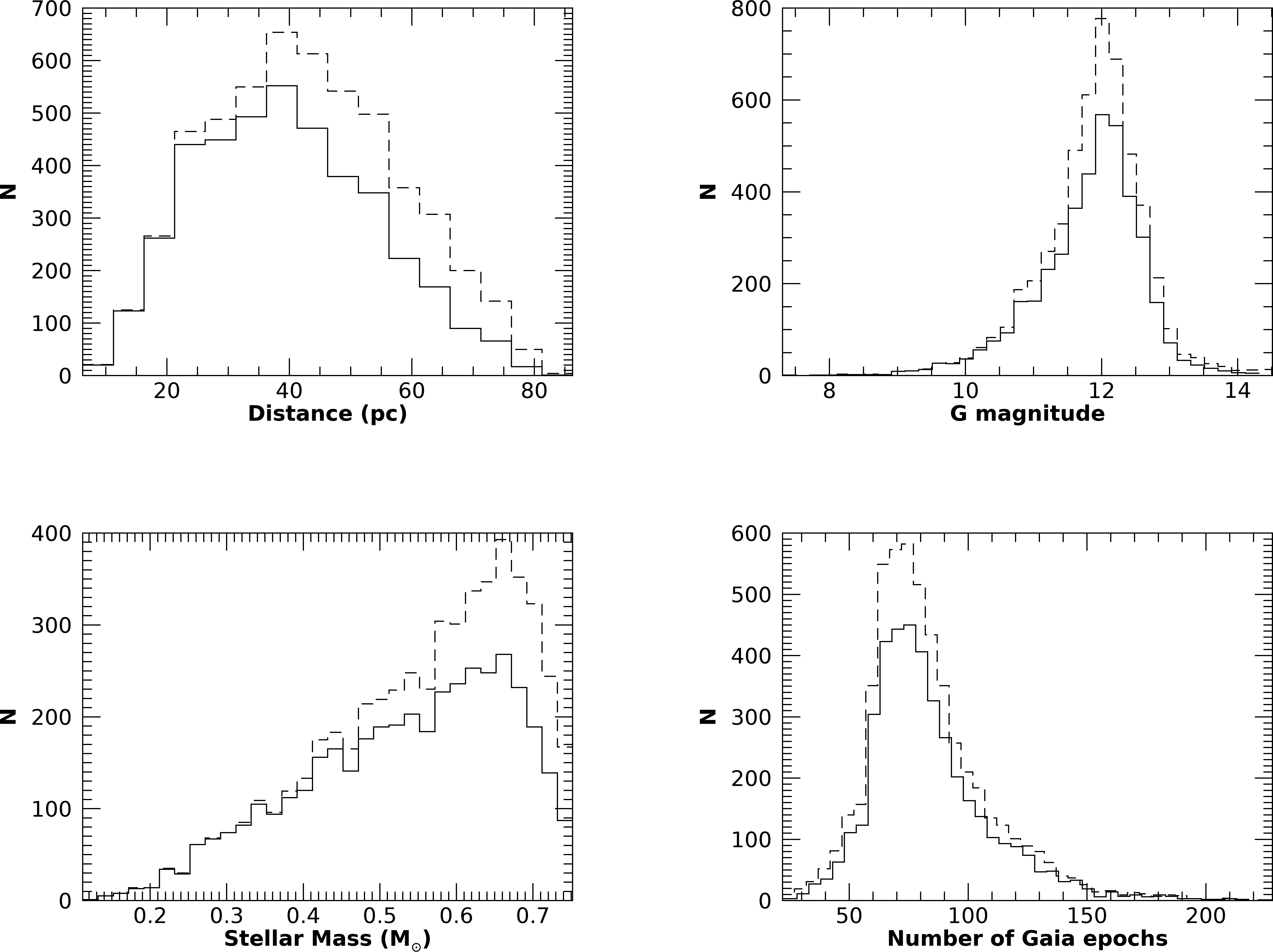} %& 
%\includegraphics[width=0.45\textwidth]{gmag} \\
%\includegraphics[width=0.45\textwidth]{mass} & 
%\includegraphics[width=0.45\textwidth]{gaiatr} \\
%\end{array} $
 \caption{Top left: distance distribution of the  M-dwarf sub-sample of the \citet{Lepine2011} successfully cross-matched to Gaia DR3. Top right: The corresponding magnitude distribution at $G$ band. 
 Bottom left: the derived distribution of primary masses. Bottom right: number of individual Gaia field transits for the same sample. See text for additional details.}
\label{fig1}
\end{figure*}

The simulation of Gaia observations follows closely the observational scenario described in \citet{Sozzetti2014}. Here we summarize and describe the main features of the setup: 

\begin{itemize}
\item[1)] The actual list of targets is based on the 8793 dM dwarf stars (in the approximate range $0.09-0.6$ $M_\odot$) from the \textit{All-sky Catalog of Bright M Dwarfs} (\citealt{Lepine2011}). 
As already noted in the Introduction, the choice of this catalogue was driven by our interest to choose a statistically significant, representative sample of relatively bright, nearby stars that would maximize detection efficiency (see Fig. 4 of \citealt{Sozzetti2014}). A coordinate-based cross-match between the \citet{Lepine2011} catalogue, the Starhorse catalogue \citep{Anders2022} and the Gaia DR3 archive returned a total of 5378 sources within $\sim100$ pc from the Sun and with masses $\lesssim0.70$ M$_\odot$. We did not investigate in detail the reasons behind the unmatched sources, as our intention is not to update the \citet{Lepine2011} catalogue, but rather select a representative sample of well-classified sources for our purposes, and the sample size returned through the cross-match exercise was deemed satisfactory. 
%Using visual and infrared magnitudes available for the sample, we utilized the color-magnitude conversion formulae of Jordi et al. (2010) to obtain $G$-band magnitudes in the Gaia broad-band photometric system. Our catalog results to have an average $G\simeq12.3$ mag. The Delfosse et al. (2000) mass-luminosity relations for low-mass stars were then utilized to obtain mass estimates for all our targets. 
In the four panels of Fig.~\ref{fig1} we show the distributions (dashed histograms) in $G$ mag, distance $d$
%\footnote{Where available, Hipparcos parallaxes are used, photometric distance estimates are otherwise utilized using the L\'epine \& Gaidos (2011) values.} 
primary mass $M_\star$, and number $N_\mathrm{obs}$ of Gaia field transits (individual along-scan measurements $w$) for our sample (a consequence of the adopted scanning law, see e.g. \citealt{Prusti2016} for details), which are those appropriate for the adopted 5-yr mission duration. The sample has median values of stellar mass, distance and magnitude are $\mathbf{M}[M_\star]\simeq0.58$ M$_\odot$, $\mathbf{M}[d]\simeq42$ pc, and $\mathbf{M}[G]\simeq11.9$, respectively. The solid histograms of Fig.~\ref{fig1} correspond to the distributions of the same parameters for the sub-sample with statistically significant orbital semi-majors axis derived in the astrometry-only fits (see Sec. \ref{sec:astrorb}). The only noticeable differences are typically slightly smaller primary masses ($\mathbf{M}[M_\star]\simeq0.55$ M$_\odot$) and slightly shorter distances ($\mathbf{M}[d]\simeq39$ pc), which is entirely expected given the scaling of the astrometric signature with $M_\star$ and $d$. 

\item[2)] The five standard astrometric parameters for each star in the sample (right ascension $\alpha$, declination $\delta$, the two proper motion components $\mu_\alpha$ and $\mu_\delta$, and the parallax $\varpi$) were taken from the Gaia DR3 archive. The generation of planetary systems proceeded as follows.
One planet was generated around each star (assumed not to be orbited by a stellar companion), with mass $M_p = 1 M_J$; orbital periods were uniformly distributed in the range $0.2\leq P\leq 5$ yr and eccentricities were distributed as a Beta function, following \citet{Kipping2013}; the orbital semi-major axis $a_p$ was determined using Kepler's third law; the inclination $i$ of the orbits was fixed to $90$ deg, 
while the two remaining angles, argument of periastron $\omega$ and longitude of the ascending node $\Omega$, where uniformly distributed in the ranges $0\leq \omega \leq 360$ deg and $0\leq \Omega \leq 180$ deg, respectively; the epoch of periastron passage was uniformly distributed in the range $0\leq T_0\leq P$. 
The resulting astrometric signature induced on the primary was calculated using the standard formula corresponding to the semi-major axis of the orbit of the primary around the barycentre of the system scaled by the distance to the observer: 
$a_\star=(M_p/M_\star)\times(a_p/d)$. With $a_p$ in au, $d$ in pc, and $M_p$ and $M_\star$ in M$_\odot$, then $a_\star$ is evaluated in arcsec.
%Furthermore we generate a second sample where the eccentricities were distributed according to the beta distribution, as a test of the last observational evidence (referenza). 

\item[3)] As in \citet{Sozzetti2014}, the error model adopts the established magnitude-dependent behavior of the along-scan formal uncertainties $\sigma_w$ at the single-transit level in Gaia astrometry, with no additional contributions from unmodeled systematics (see e.g., \citealt{Lindegren2021}, red curve in Fig. A.1). The median of the CCD-level single-measurement uncertainties is $\sim57$ $\mu$as, with a corresponding average $\sigma_w$ three times lower\footnote{we recall a full Gaia transit corresponds to 9 consecutive CCD crossings on the astrometric focal plane of the satellite}. Given the typical magnitude of the Gaia positional uncertainties involved in the simulations, the astrometric 'jitter' induced by spot distributions on the stellar surface of active dwarf stars (e.g., \citealt{Sozzetti2005,Eriksson2007,Makarov2009,Barnes2011,Sowmya2021,Meunier2020,Meunier2022}) was considered to be negligible, and  therefore not included in the error model.

\end{itemize}

\subsection{Radial velocities}

Synthetic RV datasets were produced for all cases in which the Gaia astrometric orbit has its semi-major axis determined with good statistical significance, i.e $a_\star/\sigma_a\geq5$. For all datasets for which the astrometric orbit reconstruction satisfies the above criterion, we simulate RV campaigns with time-series of 20 data points uniformly distributed over three observing seasons. With typical 6-months intervals between successive seasons, each RV follow-up campaign lasts about 2.5 yr. Uncertainties are drawn from a Gaussian distribution with standard deviation of 10 m s$^{-1}$, which is appropriate for a typical integration time of 900 sec with average observing conditions on an early- to mid-M dwarf with $V\simeq14$ mag (corresponding to $G\simeq12$ mag at the peak of the distribution of Fig. \ref{fig1}) at a 4-m class telescope equipped with a HARPS/HARPS-N-like spectrograph. These numbers are illustrative and are not necessarily optimized having in mind for example the maximization of the number of targets for follow-up within a specific amount of observing time at any given observing facility. We stress that the illustrative example of a follow-up RV campaign is designed solely to provide a sense of the amount of observing time required for refinement of the orbit of the potentially transiting gas giant at intermediate separation identified by Gaia astrometry. The rather different problem of an RV campaign that also aims at exploring, for example, the existence of low-mass companions interior to the outer massive planet is left for future work. 

\section{Models and orbit fitting algorithms}
\label{statnum}
%Here, we describe the tools utilized in the analysis of the simulated Gaia astrometric and radial velocity (RV) data.
%The analysis is done in three main step in order to reflect the observational strategy where the RV are the follow-up of the GAIA results. 

%First, statistically robust deviations from a single-star model($\psi$ depend only by the five standard astrometric parameter), indicating the presence in the astrometric observation residuals of the perturbation due to a companion with a given level of confidence, are identified through the application of a $\chi^2$-test or $F$-test (low probabilities of $P(\chi^2)$ or $P(F)$ signifying likely planet, and unlikely false positive).

The astrometry-only model (see e.g., \citealt{Holl2022} for details) adopts the following description for the time-series of one-dimensional along-scan coordinates: 

\begin{equation}\label{eq:abscissa1}
\begin{split}
%w^\mathrm{(model)} =&\, w_\mathrm{ss} + w_\mathrm{k1} \\
w^\mathrm{(model)}  =&\, ( \alpha + \mu_{\alpha} \, t ) \, \sin \psi + (\delta + \mu_\delta \, t ) \, \cos \psi + \varpi \, f_\varpi \\
 &+\, (B \, X + G \, Y) \sin \psi + (A \, X + F \, Y) \cos \psi. 
\end{split}\end{equation}

In Eq. \ref{eq:abscissa1} $f_\varpi$ and $\psi$ are the along-scan parallax factor and scan angle at time $t$, respectively, $A$, $B$, $F$, and $G$ are four of the six Thiele-Innes coefficients (e.g., \citealt{Holl2022}) and are functions of $a_\star$, $i$, $\omega$ and $\Omega$, while the elliptical rectangular coordinates $X$ and $Y$ are functions of $P$, $T_0$ and $e$. 

%\begin{eqnarray}
%\begin{split}
% \begin{array}{ll}
%A &=&  \ \ \, a_0 \; (\cos \omega \cos \Omega - \sin \omega \sin \Omega \cos i)   \\
%B &=&  \ \ \, a_0 \; (\cos \omega \sin \Omega + \sin \omega \cos \Omega \cos i)  \\
%F &=& -a_0 \; (\sin \omega \cos \Omega + \cos \omega \sin \Omega \cos i)  \\
%G &=&  -a_0 \; (\sin \omega \sin \Omega - \cos \omega \cos \Omega \cos i) 
% \end{array}
%\end{split}
%\label{eq:cu4nss_astrobin_orbital_ABFG}
%\end{eqnarray}

The combined astrometry + RV model specifically implements the representation described in \citet{Wright2009}, in which: 

\begin{equation}\label{eq:abscissa2}
%\begin{split}
%w^\mathrm{(model)} =&\, w_\mathrm{ss} + w_\mathrm{k1} \\
w^\mathrm{(model)}  = ( \alpha + \mu_{\alpha} \, t ) \, \sin \psi + (\delta + \mu_\delta \, t ) \, \cos \psi + \varpi \, f_\varpi +\, H \, S + C \, T %+ h\,\cos{\nu} + c\,\sin{\nu} + V_0.
%\end{split}
\end{equation}

\begin{equation}\label{eq:rv}
%\begin{split}
%w^\mathrm{(model)} =&\, w_\mathrm{ss} + w_\mathrm{k1} \\
RV^\mathrm{(model)}  =  \lambda\,H\,\cos{\nu} - \lambda\,C\,\sin{\nu} + V_0 = c\,\sin{\nu} + h\,\sin{\nu} + V_0.
%\end{split}
\end{equation}

In this case, $H$ and $C$ are the two remaining Thiele-Innes coefficients, $\nu$ is the true anomaly, and $V_0$ is the zero-point of the RV time-series (no provision is made to fit for long-term RV trends) while $\lambda(\varpi,P,e)$, $S(i,\Omega,X,Y,\psi)$ and $T(i,\Omega,X,Y, \psi)$ are defined as in Eqs. 67 and 74 of \citet{Wright2009}. This manipulation allows to adjust a common set of linear parameters to both astrometric and RV datasets, at the expense of now treating $i$, $\Omega$ and $\varpi$ as three additional non-linear parameters. 

%Finally, to account for potentially unmodelled signals or modelling errors, we additionally fit for a jitter term which is added in quadrature to the provided uncertainties of the along-scan abscissa, bringing the total number of fitted parameters to~13: five linear parameters for the single-source model $w_\mathrm{ss}$, seven for the single Keplerian model $w_\mathrm{k1}$ (of which the four $A$,$B$,$F$,$G$ are linear), and one non-linear jitter term $\sigma_\mathrm{jit}$.

\begin{figure*}
\centering
%$\begin{array}{cc}
\includegraphics[width=0.95\textwidth]{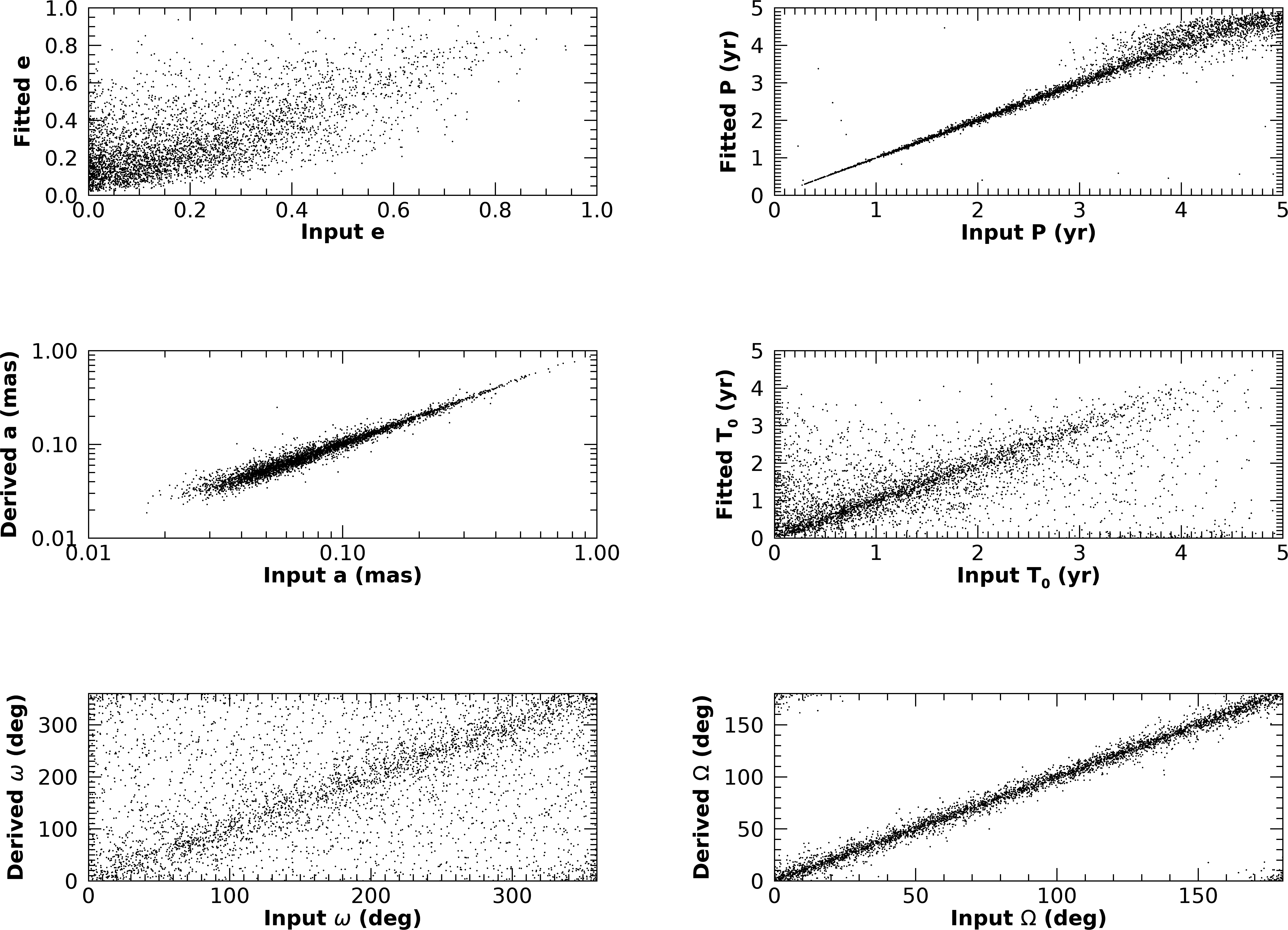}
%\end{array} $
 \caption{Fitted vs input value for orbital parameters in the astrometry-only solutions: Top left: eccentricity; top right: orbital period; center left: semi-major axis; center right: epoch of periastron; bottom left: argument of periastron; bottom right: longitude of the ascending node.
 }
\label{fig:fit_vs_true}
\end{figure*}

Orbit fitting of Gaia astrometry only and Gaia astrometry + RVs is performed using a hybrid implementation of a Bayesian analysis based on the differential evolution Markov chain Monte Carlo (DE-MCMC) method \citep{TerBraak2006,Eastman2013}. The final likelihood functions used in the DE-MCMC analysis are: 
%An earlier version of the code had been extensively tested in \citet{2008A&A...482..699C}, while its upgrade has been recently used in \citet{Drimmel2021}. In this scheme we take advantage of the four ($A$, $B$, $F$, $G$) Thiele-Innes constants representation (see Sect. \ref{ssec:mathModelDescr}. See also e.g., \citealt{Binnendijk1960,Wright2009}) to partially linearize the problem. Within this dimensionality reduction scheme, only three non-linear orbital parameters must be effectively explored using the DE-MCMC algorithm (e.g., \citealt{2008A&A...482..699C,Wright2009,Mendez2017,Drimmel2021}), namely $P$, $T_0$, and $e$. The fourth model parameter explored the DE-MCMC way is an uncorrelated astrometric jitter term $\sigma_\mathrm{jit}{}$. 
%At each step of the DE-MCMC analysis, the resulting linear system of equations is solved in terms of the Thiele-Innes constants using simple matrix algebra, QR decomposition being the method of choice. 
%\begin{equation}
%\ln \mathcal{L} = -\frac{1}{2} \left( 
%\sum_{j=1}^{{N_{astr}}}\frac{\left(w_j^{(obs)} - w_j^{(model)}\right)^2} {\sigma_{w,j}^2+\sigma_\mathrm{jit}^2}+
%\sum_{j=1}^{N_{astr}}\left(\ln\left[{\sigma_{w,j}^2 +\sigma_\mathrm{jit}^2}\right] \right )
%\right ) \label{eq:mcmcLikelihood}
%\end{equation}
\begin{align} 
-\ln \left( \mathcal{L}^\mathrm{ast} \right) \ = \ 
&\frac{1}{2} \ 
\sum_{j=1}^{{N_{ast}}}\frac{\left(w_j^{(obs)} - w_j^{(model)}\right)^2} {\sigma_{w,j}^2+\sigma_\mathrm{jit,ast}^2}+ \nonumber \\
& \frac{1}{2} \ \sum_{j=1}^{N_{ast}}\ln\left({\sigma_{w,j}^2 +\sigma_\mathrm{jit,ast}^2}\right)  \label{eq:astLnl}
\end{align}

and 

\begin{align} 
-\ln \left( \mathcal{L}^\mathrm{ast+RV} \right) \ = \ 
&\frac{1}{2} \ 
\sum_{i=1}^{{N_{ast}}}\frac{\left(w_j^{(obs)} - w_j^{(model)}\right)^2} {\sigma_{w,j}^2+\sigma_\mathrm{jit,ast}^2}+ \nonumber \\
& \frac{1}{2} \ \sum_{i=1}^{N_{ast}}\ln\left({\sigma_{w,j}^2 +\sigma_\mathrm{jit,w}^2}\right) +\nonumber \\ 
&\frac{1}{2} \ 
\sum_{j=1}^{{N_{RV}}}\frac{\left(RV_j^{(obs)} - RV_j^{(model)}\right)^2} {\sigma_{RV,j}^2+\sigma_\mathrm{jit,RV}^2}+ \nonumber \\
& \frac{1}{2} \ \sum_{j=1}^{N_{RV}}\ln\left({\sigma_{RV,j}^2 +\sigma_\mathrm{jit,RV}^2}\right) 
\label{eq:astrvlnl}
\end{align}

for the astrometry-only and astrometry+RV case, respectively. Uniform priors are used for the non-linear parameters adjusted the DE-MCMC way, the exact ranges being [0.0,5.0] yr, [$0$,$P$], [0.0,1.0], [0,$\pi$], [0,$\pi$] for $P$, $T_0$, $e$, $i$, and $\Omega$, respectively. We fit for uncorrelated jitter terms in both astrometry and RVs, which are added in quadrature to the formal uncertainties of the data, although no additional variations are included in either dataset, therefore the fitted values of $\sigma_\mathrm{jit,w}$ and $\sigma_\mathrm{jit,RV}$ are always in practice returned very close to zero, and will not be discussed further.  We refer the reader to \citet{Drimmel2021} and \citet{Holl2022} for more details on the algorithm.
%The DE-MCMC analysis is carried out with a number of chains equal to twice the number of free
%parameters. A period search is first performed in order to identify statistically more probable periodicities. Given the nature of the astrometric dataset, the direct application of publicly available tools for the periodogram analysis of unevenly sampled time-series (e.g. the Generalized Lomb-Scargle periodogram, \citealt{2009A&A...496..577Z}) is not possible. For any given source, the DE-MCMC module draws a large sample of initial trial periods for sinusoidal signals projected along the scan directions of the time-series, based on a uniform grid up to twice the observations time span. A sparsely sampled selection of periods corresponding to local $\chi^2$ minima becomes the seed for the $P$ parameter initialization of the DE-MCMC chains. Uniform priors in the ranges [$-P/2$,$P/2$] and [0,10] mas are used for $T_0$ and $\sigma_\mathrm{jit}$, respectively. Finally, starting values for $e$ are drawn from a Beta distribution, following \citet{Kipping2013}. 
%Convergence and good mixing of the chains are checked based on the Gelman–Rubin statistics (e.g., \citealt{Ford2006}). The medians of posterior distributions are taken as the final parameters. In order to comply with the choice of the main non-single-star processing chain \citep{DR3-DPACP-163}, we did not adopt the standard approach for computing 
%The $1\sigma$ formal errors on the model parameters are derived evaluating the $\pm34.13$ per cent intervals from the posterior distributions. 
Following \citet{Holl2022}, symmetric estimates of the $1\sigma$ formal uncertainties on the model parameters were obtained by reconstructing the covariance matrix directly from the Jacobians of all parameters for all observations. Standard conversion formulae (e.g., \citealt{Wright2009,Halbwachs2022}) are used to compute the Campbell elements ($a_\star$, $\omega$, and $\Omega$ and $i$, for the astrometry-only solutions, $a_\star$ and $i$ for the astrometry+RV solutions) back from the Thiele-Innes parameters, and the corresponding formal uncertainties derived using linear error propagation.

\section{Results}

\subsection{Astrometry only: quality of orbit determination}\label{sec:astrorb}

\begin{figure*}
\centering
$\begin{array}{cc}
\includegraphics[width=0.9\columnwidth]{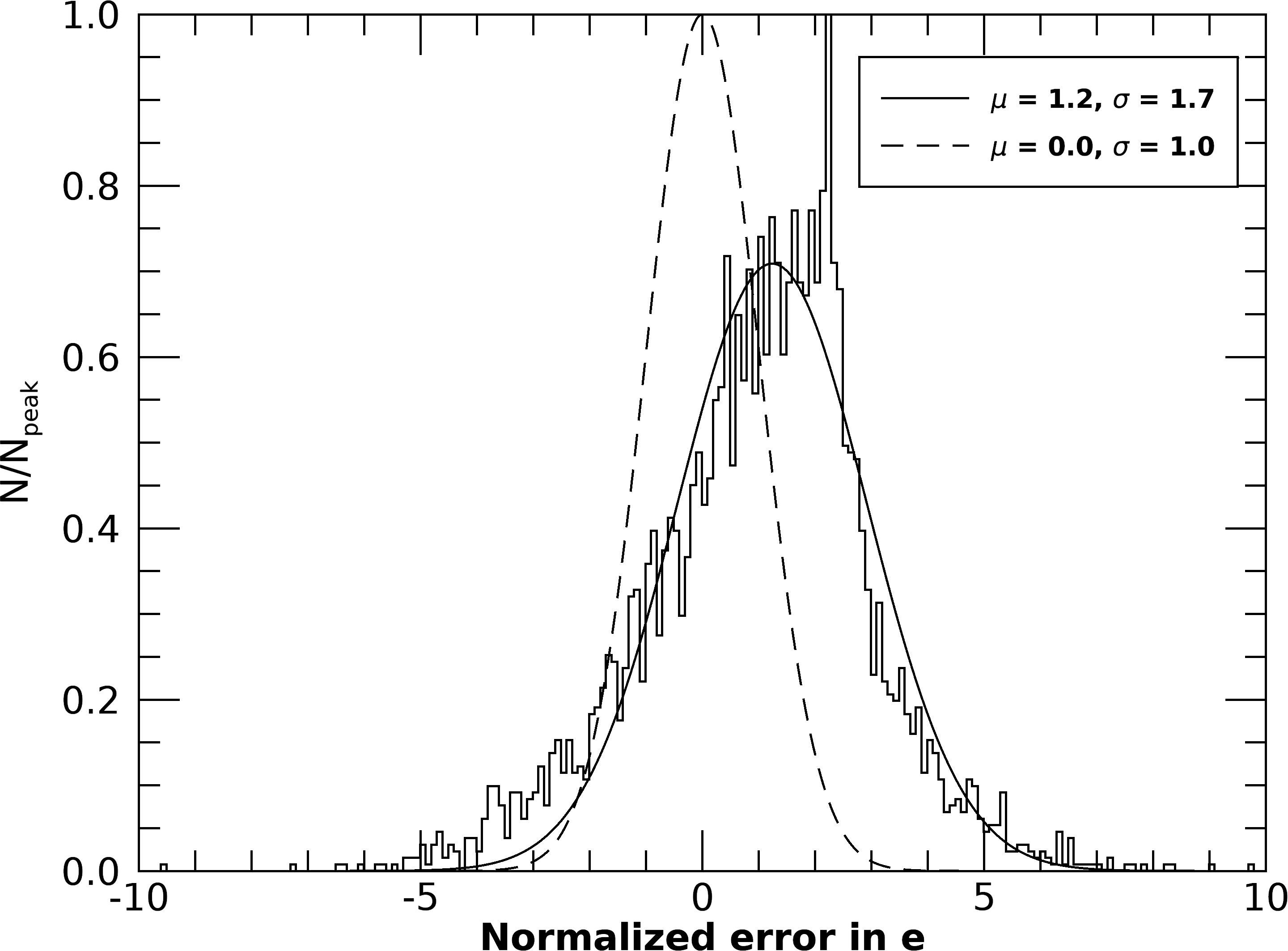} &  
\includegraphics[width=0.9\columnwidth]{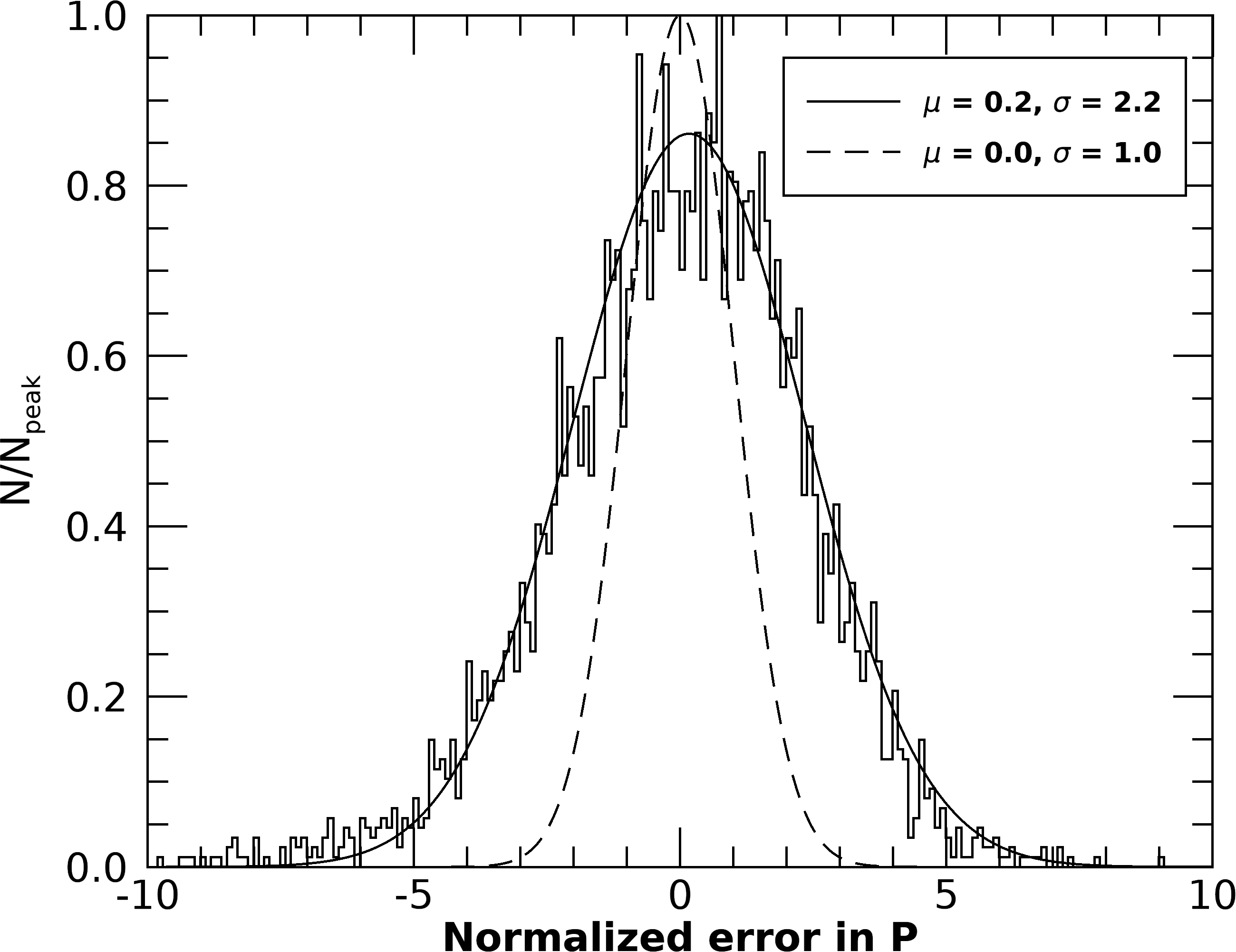} \\
\includegraphics[width=0.9\columnwidth]{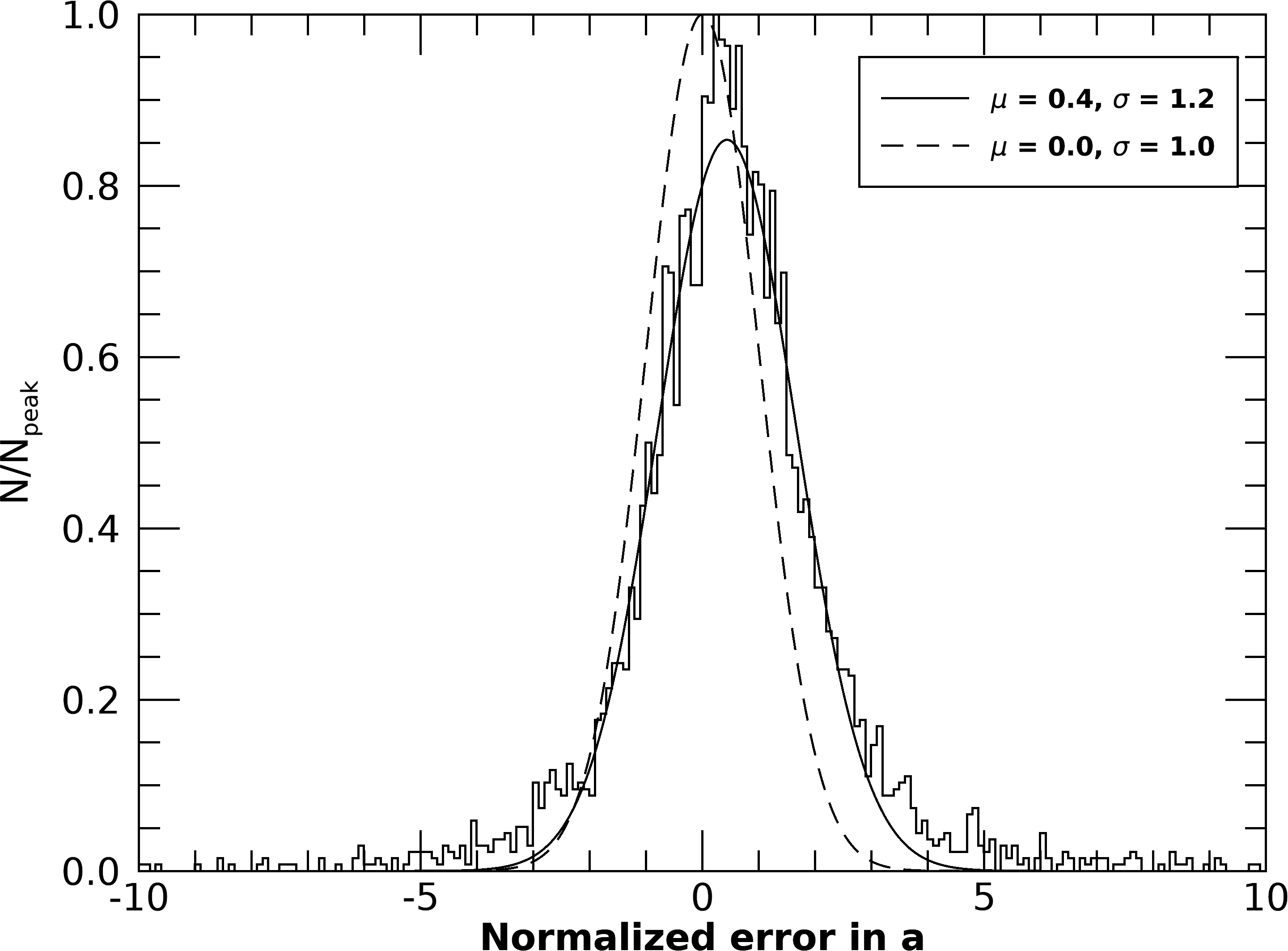} &  
\includegraphics[width=0.9\columnwidth]{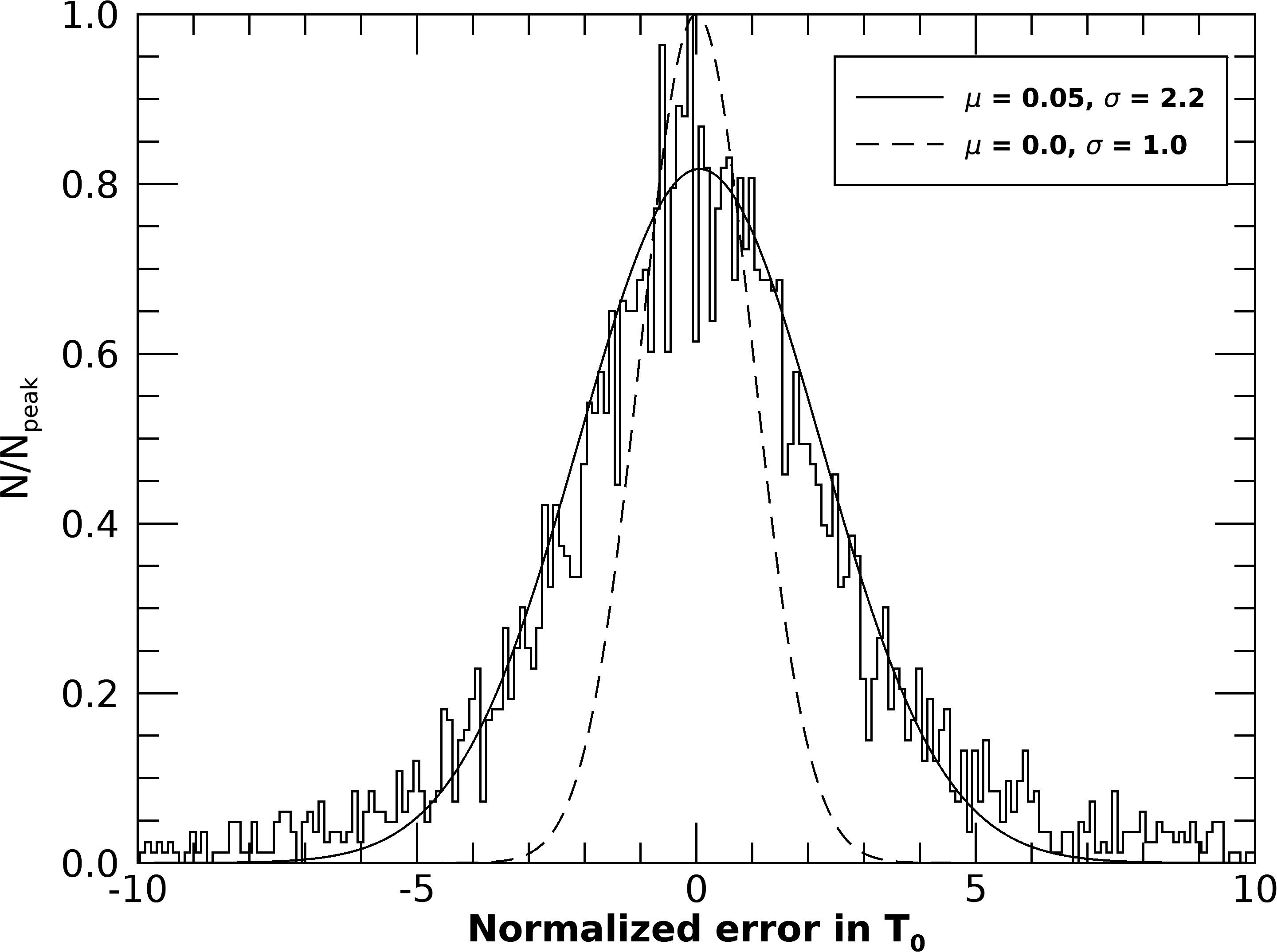} \\
\includegraphics[width=0.9\columnwidth]{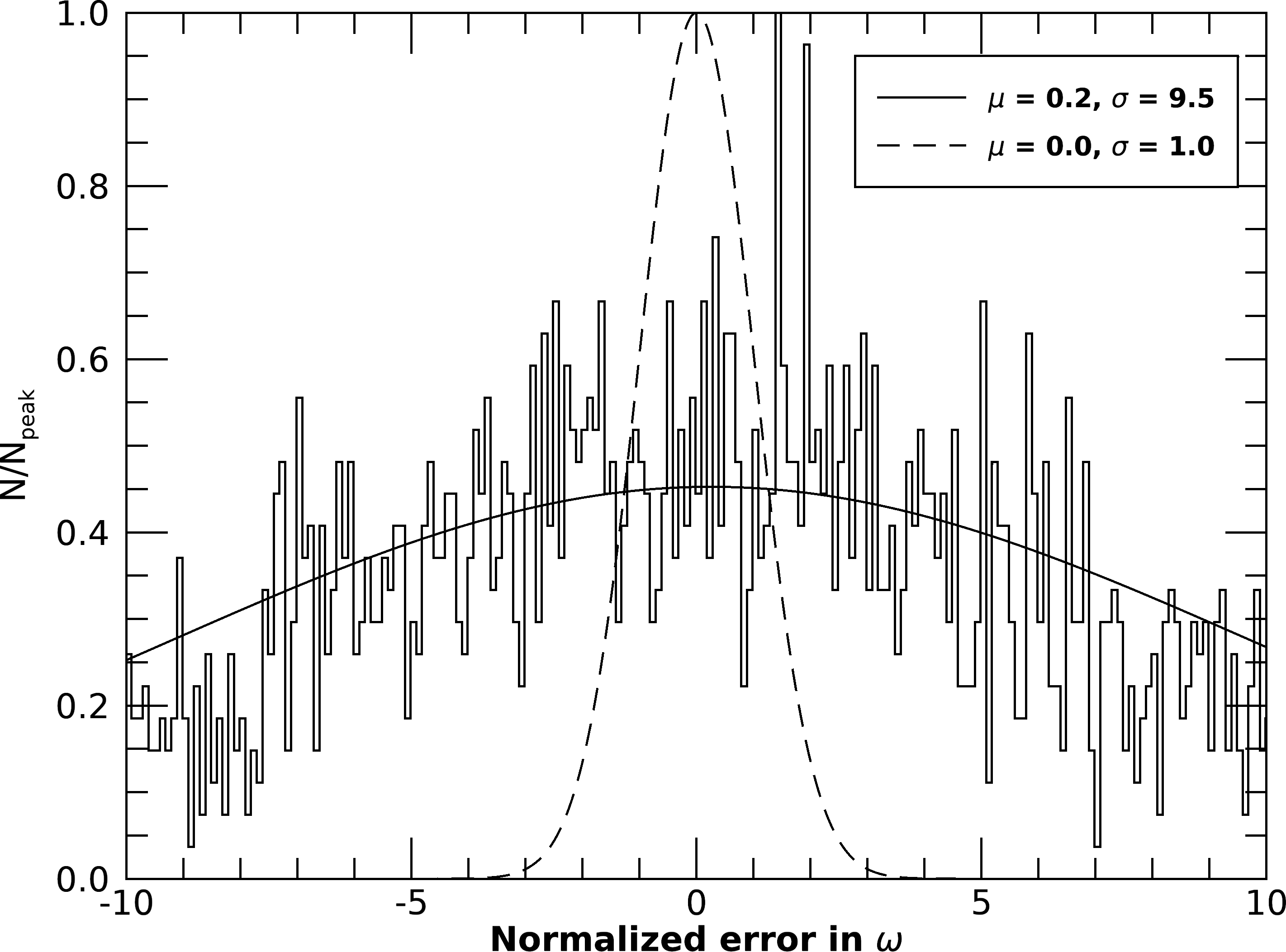} &  
\includegraphics[width=0.9\columnwidth]{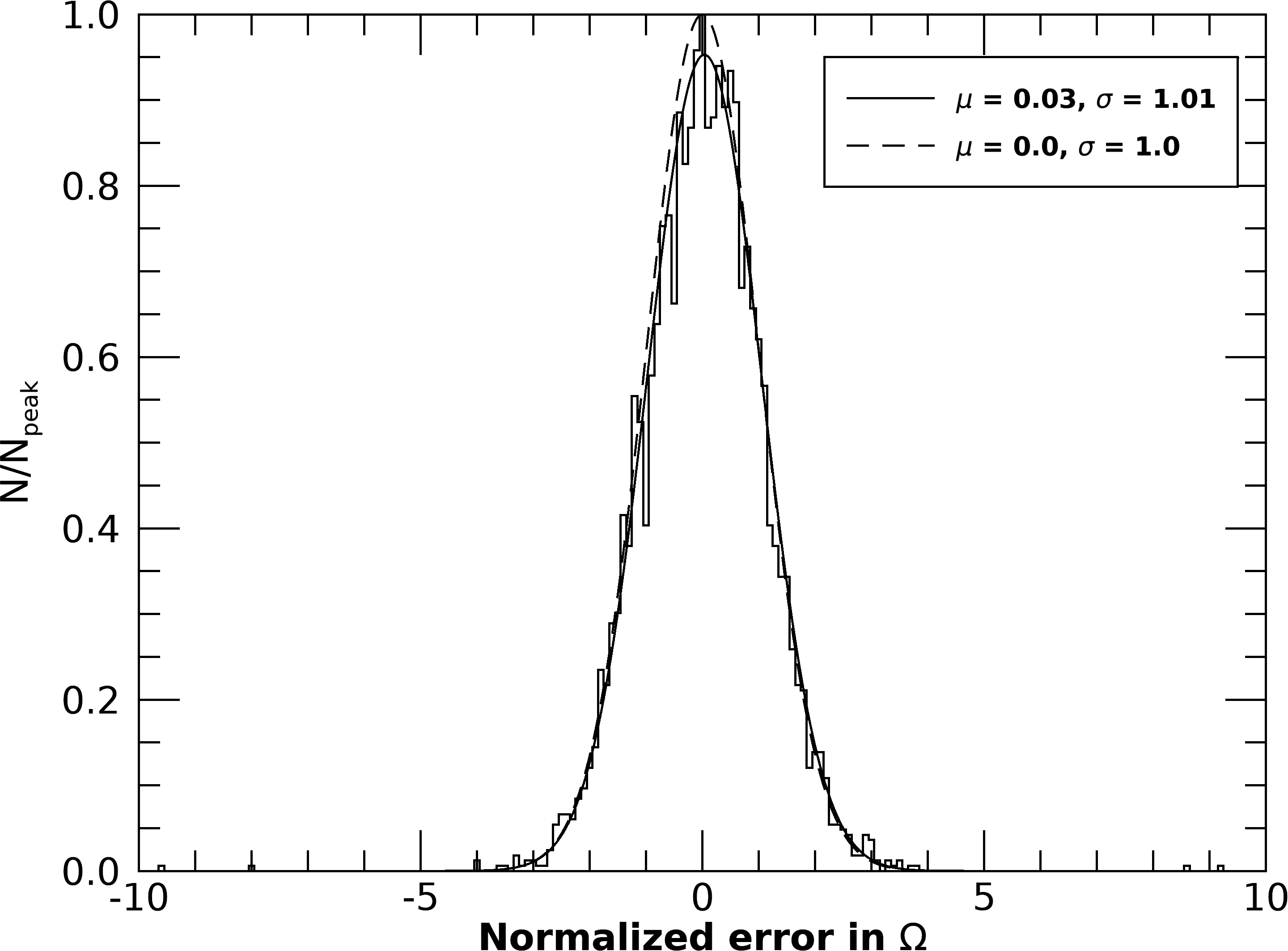} 
\end{array} $
 \caption{Distributions of normalized errors in the astrometry-only solutions (solid histograms). The ordering of the panels is the same as in Fig. \ref{fig:fit_vs_true}. The solid lines correspond to a Gaussian fit of given mean $\mu$ and standard deviation $\sigma$. The dashed lines are Gaussians of zero mean and unit dispersion.}
\label{fig:scaled_errors}
\end{figure*}

We focus our analysis on the sub-sample of 4050 astrometric orbits (75\% of the full sample) with $a_\star/\sigma_a\geq5$ for which we also performed the combination with RVs. The six panels of Figure \ref{fig:fit_vs_true} show the overall agreement between fitted or derived orbital parameters and the 'truth'. One clearly identifies a number of expected features. For example, orbital periods are accurately recovered, with only a minor loss of accuracy as $P$ approaches the nominal mission duration. Orbital eccentricity is more difficult to determine accurately, primarily due to the perfectly edge-on orbits (with consequent loss of information on the actual orbit shape), but with a second well-known effect of increasing difficulty to measure accurately values of $e\approx0.0$. The two closely related parameters $T_0$ and $\omega$ are consequently also determined with significant spread around the 1:1 correlation lines. The longitude of the ascending node is instead very well determined, while we notice a mild systematic trend of recovery of overestimated values of the angular size of the perturbation, particularly for $a\lesssim0.1$ mas. This can again be understood due to the combined effect of the edge-on configuration and the small orbit size compared to the magnitude of the individual measurement errors. 

\begin{figure}
\centering
$\begin{array}{c}
\includegraphics[width=0.9\columnwidth]{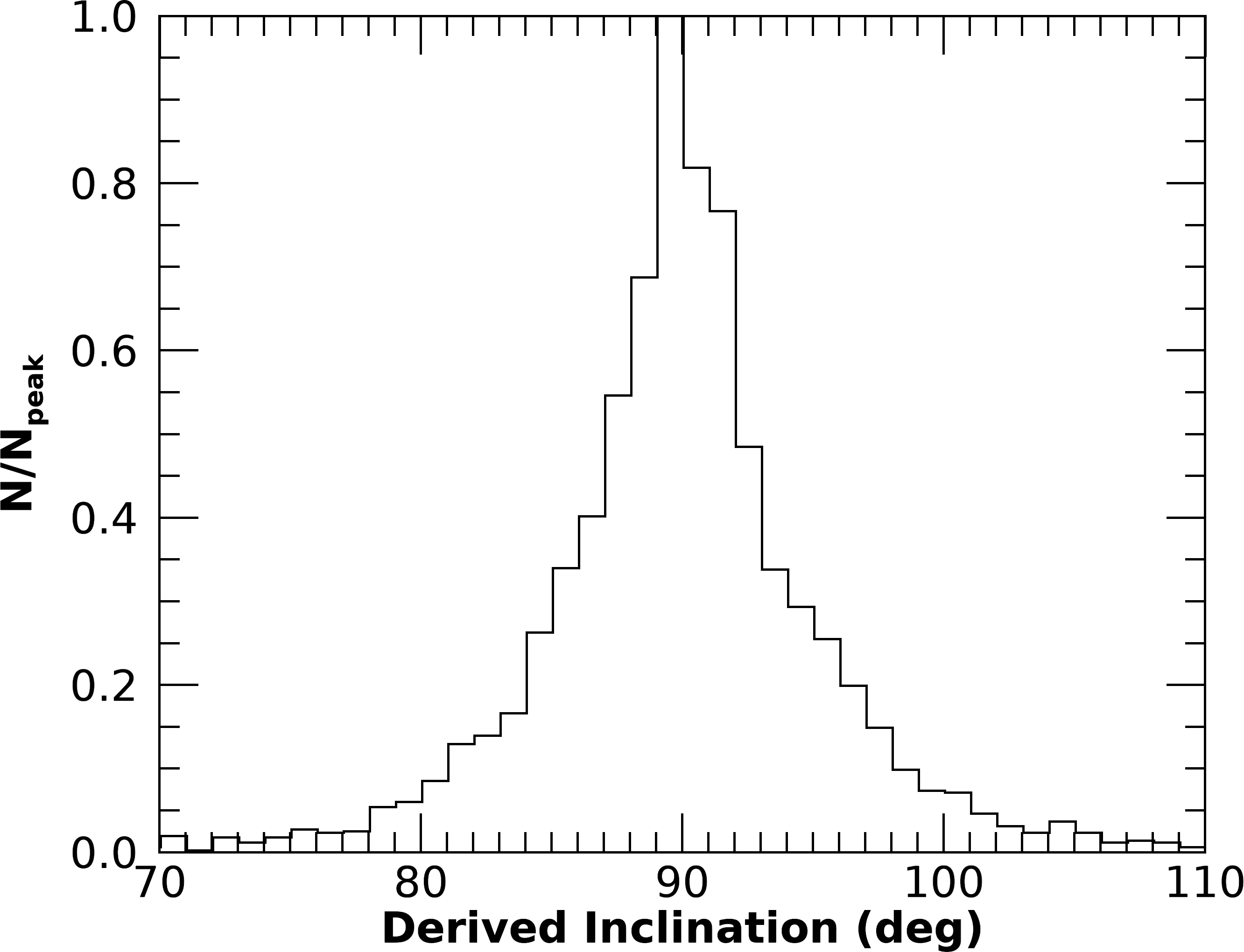} \\
\includegraphics[width=0.9\columnwidth]{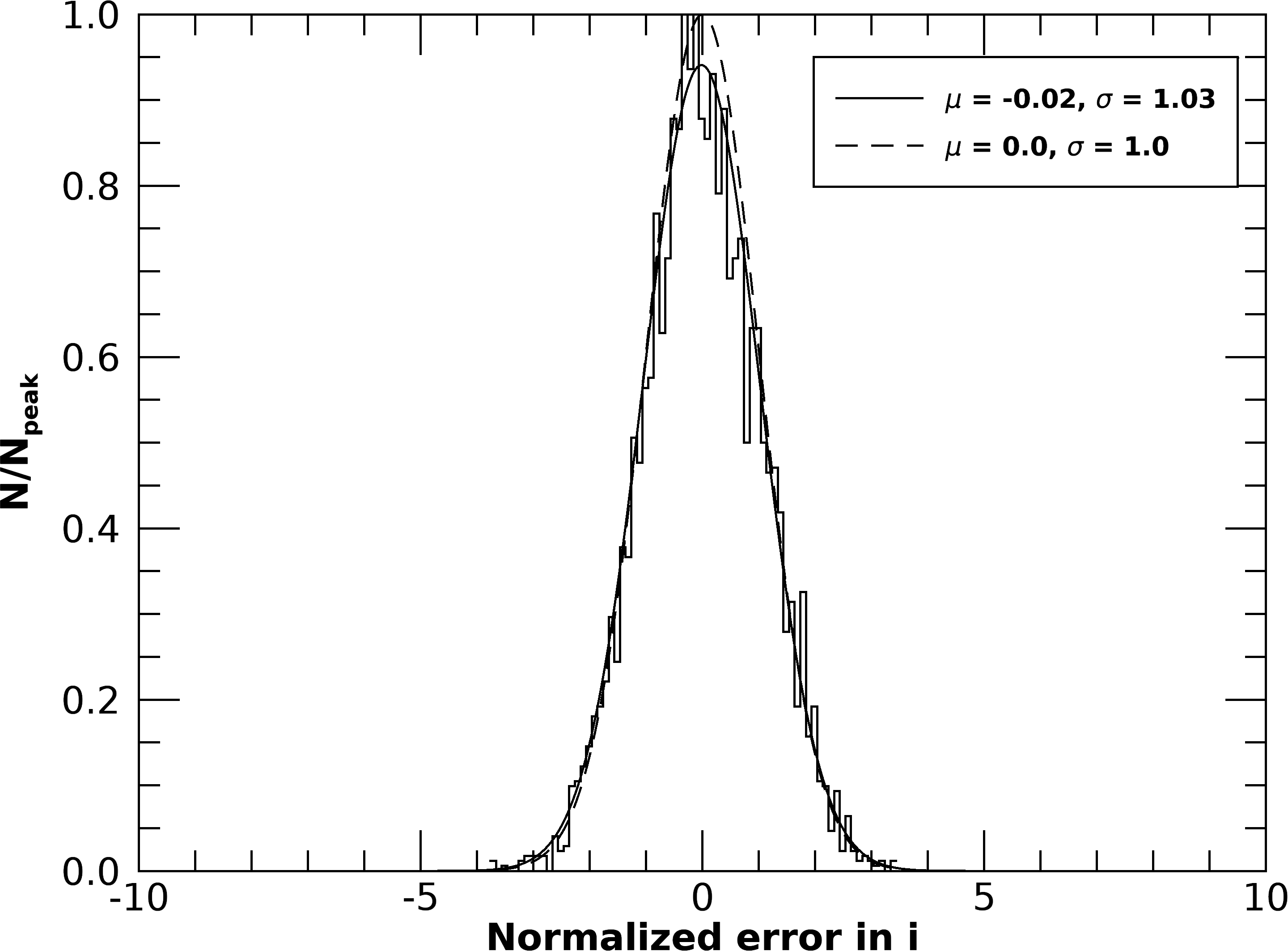} \end{array} 
$
 \caption{Top: Distribution of the derived values of orbital inclination in the astrometry-only solutions. Bottom: normalized errors in $i$. Histograms and lines have the same meaning as in Fig. \ref{fig:fit_vs_true} and \ref{fig:scaled_errors}.}
\label{fig:fitted_inc_dist}
\end{figure}

We can glean complementary insight on the quality of orbit reconstruction in our simulations using as proxy the precision with which the various orbital parameters are retrieved as a result of the orbit fitting procedure. The distribution of the ratio of the differences between the fitted and true values of a parameter divided by their estimated uncertainties (from the covariance matrix of the solution or linear propagation), which we dub here normalized errors, should be distributed normally with zero mean and unit dispersion in case the latter correctly map the former (see e.g. \citealt{Casertano2008,Holl2022}). Distortions in the distribution inform on departures from this assumption due to e.g. biases in orbit reconstruction. The six panels of Fig. \ref{fig:scaled_errors} show the distributions of normalized errors for the same parameters of Fig. \ref{fig:fit_vs_true}. The central values of the distributions for $\Omega$, $\omega$, $T_0$ and $P$ closely match the expectations. Larger positive biases for the normalized error distributions of $a$ and especially $e$ are present, as a further confirmation of the tendency to overestimate the values of the two parameters. For $\Omega$ and $a$ the width of the distribution is in excellent agreement with the expectations. For $e$, $P$, $T_0$ and $\omega$ the departure from unit dispersion gets increasingly larger, an effect understood in terms of the biases discussed for Fig. \ref{fig:fit_vs_true} for the different parameters. As formal uncertainties from the covariance matrix of the solution tend to underestimate the true errors in a significant fraction of the cases, adopting the more standard approach of evaluating the $\pm34.13$ per cent intervals from the posterior distributions should be the preferred choice. Finally, we show in the two panels of Fig. \ref{fig:fitted_inc_dist} the histogram of derived inclination angles (keeping in mind the exact $i=90^\circ$ simulated configuration) and the corresponding normalized error distribution. The the median and standard deviation of the derived inclination distribution are $89.9^\circ\pm4.5^\circ$, and 93\% of the orbits have $i$ determined to within 10 deg of a perfectly edge-on configuration. As for $\Omega$, the formal uncertainties on $i$ are found to be a very close representation of the true errors on the parameter.

\begin{figure}
\centering
%$\begin{array}{cc}
\includegraphics[width=0.99\columnwidth]{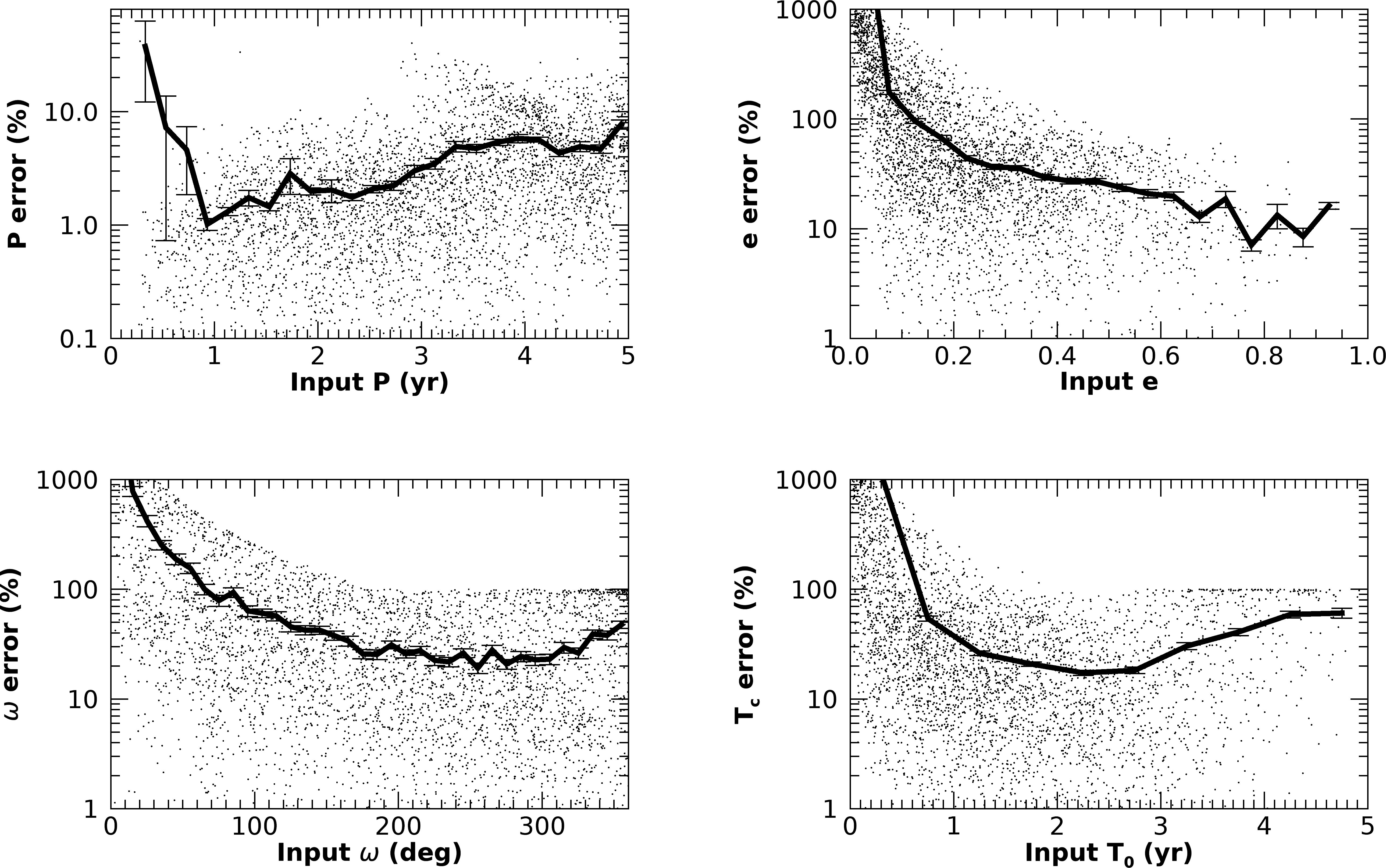}
%\end{array} $
 \caption{True errors on the relevant orbital parameters as a function of the input values, for the case of orbit determination with Gaia astrometry alone. Top left: orbital period; Top right: eccentricity; Bottom left: argument of periastron; Bottom right: epoch of periastron. In each panel, the thick solid lines correspond to the binned medians. 
 }
\label{fig:frac_errors}
\end{figure}

We next focus our attention on aspects of the quality of the determination of the orbital parameters directly affecting the forecast for the time of transit center, i.e.  $P$, $T_0$, $e$, and $\omega$ (see next Section). We discuss in particular how their fractional errors (e.g., $\sigma_P = (P_\mathrm{fitted}-P_\mathrm{true})/P_\mathrm{true}$, etc.) depend on the input values of the parameters and on the astrometric signal-to-noise ratio, defined as either $S/N_\mathrm{ast,1}=a_\star/\sigma_w$ \citep{Casertano2008} or $S/N_\mathrm{ast,2}=(a_\star/\sigma_w)*\sqrt{N_\mathrm{obs}}$ \citep{Sahlmann2015}. The four panels of Fig. \ref{fig:frac_errors} show how the fractional error on $\sigma_P$, $\sigma_{T_0}$, $\sigma_e$, and $\sigma_\omega$ varies as a function of the value of the input parameters themselves. The main features of the dependence of the precision in period determination as a function of $P$ itself were already described in e.g., \citet{Casertano2008} and \citet{Sozzetti2014}. For example, Fig.~\ref{fig:frac_errors} highlights how $\sigma_P$ increases significantly both for $P\simeq T$ as well as for short-period orbits which are under-sampled (as a direct effect of the scanning law) and translate in very low astrometric signals. On the other hand, well-sampled ($P<T$) orbital periods can be determined with $\sigma_P\sim1-2\%$, particularly in the range $1.0\lesssim P\lesssim 3.0$ yr. Similar behaviour is seen for $\sigma_{T_0}$, with an additional mild trend of improving precision with increasing eccentricity (plot not shown). The latter feature is expected as Fig. \ref{fig:frac_errors} also shows how $\sigma_e$ decreases with increasing $e$, almost circular orbits having very large fractional uncertainties. As a consequence, $\sigma_\omega$ also decreases with increasing, better-determined values of $e$ (plot not shown), and in general with increasing $\omega$ (Fig. \ref{fig:frac_errors}) . 

Overall, the median values of the fractional errors on $P$, $T_0$, $e$, and $\omega$ are $\mathbf{M}[\sigma_P]\sim2\%$, $\mathbf{M}[\sigma_{T_0}]\sim22\%$, $\mathbf{M}[\sigma_e]\sim47\%$, and $\mathbf{M}[\sigma_\omega]\sim26\%$, respectively. As expected, the quality of orbit determination also depends importantly on the astrometric signal-to-noise ratio. \citet{Casertano2008} and \citet{Sahlmann2015} had suggested thresholds of $S/N_\mathrm{ast,1}>3$ and $S/N_\mathrm{ast,2}>20$, respectively, for good astrometric orbit determination. The medians of the two diagnostics in our case are $\mathbf{M}[S/N_\mathrm{ast,1}]\sim4$ and $\mathbf{M}[S/N_\mathrm{ast,2}]\sim33$. We find that, specifically for edge-on orbit configurations, in order to improve the typical precision on the orbital parameters of interest one has to impose values of $S/N_\mathrm{ast,1}$ and $S/N_\mathrm{ast,2}$ significantly above the medians. For example, with $S/N_\mathrm{ast,1}>10$ and $S/N_\mathrm{ast,2}>60$, the median fractional uncertainties become $\mathbf{M}[\sigma_P]\sim1\%$, $\mathbf{M}[\sigma_{T_0}]\sim10\%$, $\mathbf{M}[\sigma_e]\sim18\%$, and $\mathbf{M}[\sigma_\omega]\sim11\%$,  with fractional uncertainties thus reduced by a factor of $2-3$. by further imposing $P< 3$ yr, we then obtain $\mathbf{M}[\sigma_P]\sim0.3\%$, $\mathbf{M}[\sigma_{T_0}]\sim6\%$, $\mathbf{M}[\sigma_e]\sim18\%$, and $\mathbf{M}[\sigma_\omega]\sim7\%$, with no further improvement in the typical precision in eccentricity determination. Finally, the median fractional uncertainties on $i$ is $\mathbf{M}[\sigma_i]\simeq2.8\%$, a result already derived by \citet{Sozzetti2014}. This number reduces to $\mathbf{M}[\sigma_i]\simeq0.9\%$, imposing the above mentioned thresholds on signal-to-noise ratio and orbital period.

\subsection{Astrometry only: Predicting the transit mid-time}\label{ast_tc}

\begin{figure*}
\centering
%$\begin{array}{cc}
\includegraphics[width=0.95\textwidth]{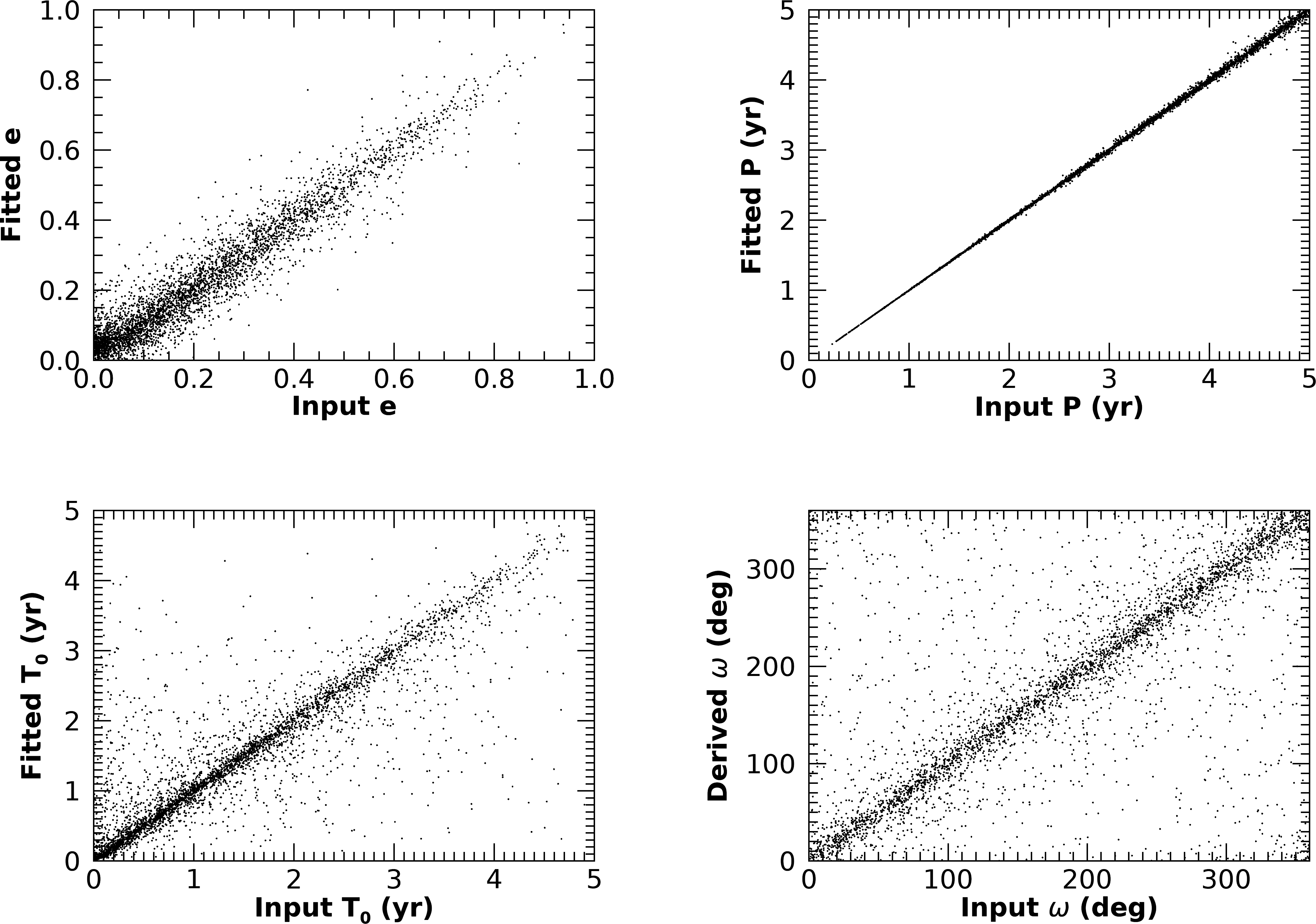}
%\end{array} $
 \caption{Fitted vs input value for the four orbital parameters relevant for the determination of the time of transit center in the combined astrometry+RV solutions. Top left: eccentricity; top right: orbital period; bottom left: epoch of periastron; bottom right: argument of periastron. 
 }
\label{fig:fit_vs_true_astro_rv}
\end{figure*}

The time of transit center $T_c$ can be computed given $P$, $T_0$, $e$ and $\omega$ as follows (e.g., \citealt{Irwin2008}): 

\begin{equation}
    T_c = T_0 + \frac{P}{2\pi}\left(E-e\sin E\right),
\end{equation}

where the eccentric anomaly $E=2\arctan[(1-e)(1+e)]^{1/2}\tan(\nu/2)$, and with the true anomaly at the time of central transit calculated as $\nu=\pi/2-\omega$ for a perfectly edge-on orbit. The fractional uncertainty $\sigma_\mathrm{T_c}$ as a function of the input $T_c$ value (Figure not shown) exhibits a clear similarity with the plot of $\sigma_\mathrm{T_0}$ vs $T_0$, underlining the fact that the uncertainties in the determination $P$, $e$ and $\omega$ play only a minor role in forecasting the time of transit center. Indeed, the median value $\mathbf{M}[\sigma_{T_c}]\sim21\%$, virtually identical to that of $\sigma_\mathrm{T_0}$. In units of days, the median uncertainty for the full sample is $\sigma_\mathrm{T_c}\sim97$ d. This number decreases by a factor $\sim2$ if we consider either $P<3$ yr or $S/N_\mathrm{ast,1}>10$ ($S/N_\mathrm{ast,2}>60$), while $\sigma_\mathrm{T_c}\sim20$ d in the case of the sub-sample satisfying both conditions simultaneously. The intrinsic $\pm\pi$ ambiguity in the determination of the argument of periastron when only astrometric measurements are available \citep{Wright2009} however implies that in practice two $T_c$ values will need to be predicted, i.e. for $\omega+\nu=90^\circ$ or $270^\circ$, with a corresponding doubled need of observing time investment for any follow-up programs probing the actual transiting nature of the detected companions \citep{Perryman2014}.

\subsection{Combining Astrometry with Radial Velocities: orbit determination and transit mid-time forecast}\label{astrrvorb}

As described in Sec.\ref{statnum}, RV follow-up campaigns were simulated in case of a $ \geq 5-\sigma$ significance of the orbital semi-major axis. We focus here on quantifying the improvements in the determination of the orbital parameters directly affecting the forecast of the transit window. As is clearly seen in Fig. \ref{fig:fit_vs_true_astro_rv}, there is a much tighter agreement between the fitted/derived values of $P$, $e$, $T_0$ and $\omega$ with respect to the astrometry-only case. The significant gain with the combined fits is quantified looking at the fractional error distributions for $P$, $T_0$, $e$, and $\omega$ discussed in Sec. \ref{sec:astrorb}: these are significantly reduced, with median values $\mathbf{M}[\sigma_P]\sim0.4\%$, $\mathbf{M}[\sigma_{T_0}]\sim8\%$, $\mathbf{M}[\sigma_e]\sim17\%$, and $\mathbf{M}[\sigma_\omega]\sim8\%$, respectively. With the additional constraints on signal-to-noise ratio and orbital period discussed before, the above numbers become $\mathbf{M}[\sigma_P]\sim0.1\%$, $\mathbf{M}[\sigma_{T_0}]\sim3\%$, $\mathbf{M}[\sigma_e]\sim7\%$, and $\mathbf{M}[\sigma_\omega]\sim4\%$, respectively, with uncertainties reduced by factors $2.0-4.0$. As expected, the quality of determination of the inclination angle is instead unaffected by the addition of RV measurements (at least for the illustrative examples of RV campaign utilized in this work), with median fractional errors essentially identical to the ones reported in the previous Section.

In terms of forecasting the time of transit center, the median value $\mathbf{M}[\sigma_\mathrm{T_c}]\sim8\%$, or $\sim34$ d. This number decreases by factors of $\sim1.4$, $\sim2.3$ and $\sim4.2$ if we consider the sample with $S/N_\mathrm{ast,1}>10$ ($S/N_\mathrm{ast,2}>60$), $P<3$ yr, and with both constraints applied, respectively. 

In the simplest case of a central transit and circular orbit, the transit duration is $t_d\approx3\,\mathrm{hr}\,\left(\frac{P}{4\mathrm{d}}\right)^{1/3}\,\left(\frac{\varrho_\star}{\varrho_\odot}\right)^{-1/3}$ \citep{Seager2003}. For $P=2$ yr and a typical $\varrho_\star\simeq5.5\varrho_\odot$ for the stellar sample under consideration here, then $t_d\simeq10$ hr. The uncertainty on the transit mid-time therefore remains the dominant factor even in the most favorable cases, with typical transit windows of $\sim\pm2$ weeks at the $\sim2\sigma$ level. 

Finally, we also experimented doubling the amount of RV data within the same duration of the observing campaign. We recovered the expected improvement of a factor $\simeq\sqrt{2}$ in the median precision with which $P$, $e$, $T_0$, $\omega$, and therefore $T_c$, were determined for the full sample.

\begin{table}%[ht!]
    \centering
%	\small
	\caption{Median uncertainties on $T_c$ based on orbit determination with Gaia astrometry alone and with the combined of Gaia astrometry and ground-based RVs observations.	\label{tab:summary}
}
	\begin{tabular}{lcc}
    \hline
    \noalign{\smallskip}
    Case     &  Astrometry & Astrometry\\
         &  only & +RV \\
%         &  (full sample) &  ($S/N_\mathrm{ast,1} > 10$) & ($P< 3$ yr) & ($S/N_\mathrm{ast,1} > 10$) \& ($P< 3$ yr) \\
    \noalign{\smallskip}
    \hline
    \noalign{\smallskip}
    \noalign{\smallskip}
%    \textit{Astrometric:} \\
    \noalign{\smallskip}
    $\mathbf{M}[\sigma_{T_c}]$ (d), full sample & 96.9 & 34.1  \\
    \noalign{\smallskip}
    $\mathbf{M}[\sigma_{T_c}]$ (d), $S/N_\mathrm{ast,1} > 10$ & 47.3 & 24.3 \\
    \noalign{\smallskip}
    $\mathbf{M}[\sigma_{T_c}]$ (d), $P< 3$ yr & 45.6 & 14.7  \\
    \noalign{\smallskip}
    $\mathbf{M}[\sigma_{T_c}]$ (d), $S/N_\mathrm{ast,1} > 10$ \& $P< 3$ & 19.7 & 8.4  \\
    \noalign{\smallskip}
    \hline
    \end{tabular}
%    \\
%    \tablefoot{
%    \tablefoottext{a}{From spectral analysis.}
%    \tablefoottext{b}{Trigonometric surface gravity using Gaia EDR3 data.}}
%    \\
%    \tablebib{[1] \cite{2020yCat.1350....0G}; [2] \cite{2012yCat.1322....0Z}; [3] \cite{2003yCat.2246....0C}; [4] This work.}
\end{table}

\section{Summary and Discussion}

In this paper we have gauged how the class of transiting cold Jupiters uncovered astrometrically by Gaia would benefit from the availability of additional Doppler measurements aimed at improving the accuracy of the orbital solutions and the corresponding transit ephemeris predictions for the purpose of confirming or ruling out the fact that the Gaia-detected companions do indeed transit. Our main findings can be summarized as follows: 

\begin{itemize}
    \item Based on realistic simulations of Gaia observations of systems composed of Jupiter-mass companions around a sample of nearby low-mass stars and state-of-the-art orbit fitting tools, we have shown how forecasts of the time of transit center will carry typical uncertainties of a few months, which might be reduced to about three weeks in the limit of very high astrometric signal-to-noise ratio and orbital periods $<3$ yr; 
    \item We have implemented a framework for combined astrometry+radial velocity orbital fits, and gauged the benefits of illustrative RV campaigns towards significant improvements in the identification of the possible transit windows, which will see typically reductions of the two figures above by a factor $\sim3$, with the added bonus of resolving the $\pm\pi$ ambiguity in the determination of the argument of periastron. A further summary of the key results is provided in Table \ref{tab:summary}. 

\end{itemize}

\citet{Sozzetti2014}, using the Besancon galaxy model of stellar populations \citep{Robin2003} and reasonable assumptions for the frequency of giant planets within 3 au around M0-M9 dwarf primaries at $d\leq100$ pc, estimated that, for fractional uncertainties on the inclination angle of 10\%, 5\% and 2\%, Gaia alone could detect 255, 85 and 10 systems, respectively, formally compatible with transiting configurations within the $1\sigma$ error bars. \citet{Perryman2014} provided a figure of merit of $\sim650$ detectable giant planets with $1<P<10$ yr and $|\cos i| <0.1$ around F-G-K-M dwarfs out to $\sim400-500$ pc. The typical 2.5-yr RV follow-up campaign of transit candidates within $P\lesssim5$ yr would entail an investment of $\sim0.5$ observing nights per target at 4-m class facility equipped with a HARPS/HARPS-N like instrument. Monitoring of the top 100 candidates with the best constraints on orbital inclination would therefore require $\sim50$ nights of observing time distributed over four observing semesters. This ballpark estimate indicates systematic RV follow-up campaigns of Gaia astrometrically detected candidate transiting gas giants are definitely feasible investing reasonable amounts of observing time. A potentially relevant caveat will however concern the achievement of the optimal balance in size of the candidate sample for follow-up based on updated expectations of false positive rates, which will better gauged when robust Gaia survey sensitivity estimates will become available. 

In principle, follow-up photometric observations for confirmation of the transiting nature of the giant planetary companions could be carried out from the ground even with modest-size telescopes. However, while the transit depths ($\sim1\%$) will have a magnitude readily accessible with ground-based facilities, the typical duration of the events will exceed that of a full observing night, requiring challenging multi-site campaigns for detection of partial events, such as the ones recently carried out to capture the transits of the long-period giant planets HD 80606 b \citep{Pearson2022}, HIP 41378 f \citep{Bryant2021}, and Kepler-167 e \citep{Perrocheau2022}, with orbital periods in the approximate range $100-1000$ days\footnote{Provided the primaries are bright and not slow rotators, a similar approach using multiple facilities for high-precision RV work can be implemented to follow-up long-period, lower-mass companions for the purpose of measurement of the Rossiter-McLaughlin effect, as it was successfully demonstrated recently in the case of HIP 41378 d \citep{Grouffal2022}.}. In addition, opportunities for follow-up from the ground will be very rare. As already suggested by \citet{Sozzetti2014} and \citet{Perryman2014}, it might be beneficial to revisit photometric light-curve databases of long-term ground-based transit programs (e.g., Super-WASP, HATNet, HATSouth, MEarth, APACHE, etc.), looking for missed or uncategorized transit events in the time-series of the candidates (e.g., \citealt{Cooke2018,Kovacs2019,Yao2019,Yao2021}). It will be however space-based transit photometry the likely most effective provider of confirmation/refutation measurements. Any candidates from Gaia in the original Kepler field would immediately benefit from the 4-yr long, continued Kepler photometry. The K2 mission and the TESS extended mission might also contribute to the task. Continuous, multi-year photometric monitoring of the fields that will ultimately be selected in the planned combination of long-duration observation and step-and-stare phases of the PLATO mission \citep{Nascimbeni2022} will eventually be a crucial source of follow-up measurements of Gaia transiting planet candidates over a large fraction ($\sim40\%$) of the observable sky. Our results reinforce the notion that Gaia astrometric detections of potentially transiting cold giant planets around bright stars, starting with Data Release 4, will constitute a valuable sample worthy of synergistic follow-up efforts with a variety of techniques, to identify those for which it might be possible in practice to perform spectroscopic characterization of their atmospheres.

\section*{Acknowledgements}

This work has made use of data from the European Space Agency (ESA) mission {\it Gaia} (\url{https://www.cosmos.esa.int/gaia}), processed by the {\it Gaia} Data Processing and Analysis Consortium (DPAC,
\url{https://www.cosmos.esa.int/web/gaia/dpac/consortium}). Funding for the DPAC has been provided by national institutions, in particular the institutions participating in the {\it Gaia} Multilateral Agreement. We acknowledge financial contribution from the agreement ASI-INAF n.2018-16-HH.0. We gratefully acknowledge support from the Italian Space Agency (ASI) under contract 2018-24-HH.0 "The Italian participation to the Gaia Data Processing and Analysis Consortium (DPAC)" in collaboration with the Italian National Institute of Astrophysics.

\section*{Data Availability}

The stellar data used in this study are available through the Gaia archive facility at ESA (https://gea.esac.esa.int/archive/) and the All-sky Catalog of Bright M dwarfs database (https://vizier.cds.unistra.fr/viz-bin/VizieR-3?-source=J/AJ/142/138). The synthetic data (simulation results) underlying this publication will be shared on reasonable request to the corresponding author.



\bibliographystyle{mnras}
\bibliography{Gaia_transit} % if your bibtex file is called example.bib

%\begin{thebibliography}{99}
%\end{thebibliography}

%%%%%%%%%%%%%%%%%%%%%%%%%%%%%%%%%%%%%%%%%%%%%%%%%%

%%%%%%%%%%%%%%%%% APPENDICES %%%%%%%%%%%%%%%%%%%%%

%\appendix



%%%%%%%%%%%%%%%%%%%%%%%%%%%%%%%%%%%%%%%%%%%%%%%%%%

% Don't change these lines
\bsp	% typesetting comment
\label{lastpage}
\end{document}